\documentclass[%
 reprint,
showpacs,
 amsmath,amssymb,
 aps,
]{revtex4-1}

\usepackage{graphicx}
\usepackage{dcolumn}
\usepackage{bm}
\usepackage{gensymb}
\usepackage[justification=raggedright]{caption}
\usepackage{subfig}
\usepackage{color}
\usepackage{hyperref}

\begin{document}

\preprint{APS/123-QED}

\title{Effects of magnetic field topology in black hole-neutron star mergers:\\Long-term simulations}

\author{Mew-Bing Wan}
\affiliation{(a) Institute for Advanced Physics and Mathematics, Zhejiang University of Technology, Hangzhou, 310032, China}
\affiliation{(b) Asia-Pacific Center for Theoretical Physics, POSTECH, Pohang 37673, South Korea}
\affiliation{(c) Center for Theoretical Physics of Complex Systems, Institute for Basic Science, Daejeon 34051, South Korea}
\affiliation{(d) Korea Astronomy and Space Science Institute, Daejeon 34055, South Korea}

\date{\today}

\begin{abstract}
We report long-term simulations of black hole-neutron star binary mergers
where the neutron star possesses an asymmetric magnetic field dipole.
Focusing on the scenario where the neutron star is tidally disrupted by the black hole,
we track the evolution of the binary up to $\approx 100$ms after the merger. 
We uncover more than one episode of thermally driven winds being launched along a funnel wall in all these cases beginning from $\approx 25$ms after the merger.
On the other hand, we are unable to conclude presently whether the amount of ejected mass increases with the \textit{degree} of asymmetry. 
A large-scale magnetic field configuration in the poloidal direction is formed along the funnel wall accompanied by the generation of a large Poynting flux.
The magnetic field in the accretion disk around the black hole remnant is amplified by both magnetic winding and the nonaxisymmetric magnetorotational instability (MRI).
The MRI growth is estimated to be in the ideal magnetohydrodynamics (MHD) regime and thus would be free from significant effects induced by potential neutrino radiation.
However, the asymmetry in the magnetic field leads to increased turbulence, which causes the vertical magnetic field in the accretion disk to grow largely in a nonlinear manner.

\end{abstract}

\maketitle


\section{\label{sec:level1}Introduction}

The question of whether a short gamma-ray burst (sGRB) is generated by a compact binary merger is still under active investigation. 
From the observational front, the location and distribution of sGRBs in elliptical galaxies with large offsets from their hosts, together with the lack of an association with star-forming regions and supernovae, have thus far lent indirect support for the hypothesis \cite{Nakar07,Berger14}.
A strong direct potential evidence was proposed by Li and Paczynski in 1998 \cite{Li98} in the form of a kilonova emitted via the radioactive decay of $r$-process material ejected by a compact binary merger.
A possible kilonova associated with the sGRB 130603B was observed in 2013 by Tanvir, \textit{et al} \cite{Tanvir13} as a transient in the near infrared band in line with the earlier possible detection by Berger, \textit{et al} \cite{Berger13}.
Subsequently, a reanalysis of the GRB 060614 optical afterglow data have yielded the first ever multiepoch/band light curves that have considerable agreement with black hole-neutron star (BHNS) merger kilonovae \cite{Yang15,Jin15,Jin16}.
These findings reinforced the compact binary progenitor hypothesis as a frontrunner theory of sGRB generation. 

Incidentally, compact binaries have also been identified as the most promising sources of gravitational waves, and the first ever detections pointing to binary black hole (BBH) systems as sources, have been announced in the GW150914 and GW151226 events by the Laser Interferometer Gravitational-Wave Observatory (LIGO) Scientific Collaboration and the Virgo Collaboration (LVC) \cite{Abbott16,Abbott16a}.
Given the predicted rates and sensitivity of the Advanced LIGO/VIRGO/KAGRA detectors \cite{Abadie10,Abbott16b}, the detection of gravitational waves from compact binary mergers involving at least one neutron star (NS), is imminent.
A sGRB would serve as an excellent electromagnetic counterpart to gravitational waves from these compact binary mergers.

The sGRB engine in this picture is comprised of a hot, dense and massive accretion disk around a remnant rotating black hole (BH) formed after a compact binary merger \cite{Nakar07}.
The lifetime of the disk is comparable to the duration of the sGRB, implying hyperaccretion rates several orders of magnitude greater than those of other accreting BH systems such as active galactic nuclei and microquasars \cite{Nakar07}.
In BHNS mergers, it has been shown in fully general relativistic hydrodynamics simulations that the disk is formed when the NS is tidally disrupted by the companion BH \cite{Shibata11,Kyutoku13,Foucart12,Foucart13,Lovelace13}. 
The launch of the relativistic jet that results in the sGRB is then predicated upon the presence of an initially baryon-poor region along the BH rotation axis \cite{Uso92,Price06,Nakar07}, where matter outflow is highly collimated with high Lorentz factors. 
A variety of studies on these jets have unravelled several key factors on how they could emerge in BHNS mergers (for an early effort in NSNS mergers, see Ref.~\cite{Rezzolla11}), and we review these as follows in order to lay the foundations that motivate the current study.

Numerical simulations using general relativistic magnetohydrodynamics (GRMHD) with a fixed background space-time, have found that for effective collimation, the presence of a large-scale poloidal magnetic field configuration is essential \cite{Beckwith08}. This is corroborated by full GRMHD+force-free electrodynamics (FFE) simulations in Ref.~\cite{Paschalidis15} where the inspiralling NS is endowed with a magnetosphere instead of an embedded magnetic field dipole, producing a jet in the aftermath of the merger, basically hinging on the action of the magnetosphere in the magnetically dominant regions outside the NS. 
In the case where the field in the accretion disk are predominantly toroidal and frozen into highly conducting matter, could BHNS mergers generate a large-scale poloidal field in the merger aftermath? 
This question was addressed recently by full GRMHD simulations in Ref.~\cite{Kiuchi15}, which point to the production of a thermally driven torus wind as the agent generating the large-scale poloidal field in the merger remnant. 

With these studies shedding light on how large-scale fields play a crucial role in jet formation and how they could be generated, we consider next the field strengths. Reference~\cite{Balbus92} addressed this issue when a powerful local instability is found to be ubiquitous in accretion disks. This instability, called the magnetorotational instability (MRI), exponentially amplifies even the smallest magnetic field, while facilitating rapid angular momentum transport and turbulence. 
The maximum growth rate for the MRI is captured in the axisymmetric limit of a nonaxisymmetric system, where the wave numbers of magnetic field perturbations in the vertical direction greatly exceed that in the toroidal direction \cite{Balbus98,Hawley11}. This means that having a wisp of a vertical field in the system dramatically enhances the MRI growth rate. 

A combination of an amplification mechanism and a mechanism of large-scale generation in the magnetic field seems imperative for the jet formation necesssary for a sGRB.
The study in Ref.~\cite{Kiuchi15} realized high effective turbulent viscosities via high-resolution simulations, which enhances the rate of thermalization of the accreting mass.
This in turn generates the wind structure that drags the magnetic field into a large-scale configuration along a funnel wall. 
Could there be other degrees of freedom found naturally in realistic astrophysical systems which could enhance the prospects of jet formation by increasing the strengths of these large-scale field?
Full GRMHD simulations in Ref.~\cite{Etienne12} have pointed out that asymmetry plays a role in enhancing poloidal fields.
Reference~\cite{Thompson93} has proposed that, due to convection inside the core, the direction of the dipole in a pulsar, for example, would not be aligned with its rotation axis. 
In Ref.~\cite{Braithwaite04}, it has been shown via numerical simulations that random magnetic fields in a self-gravitating body almost always settle down into a long-term stable helical configuration.
This was suggested as an indication that the long-term stable strong magnetic fields inside magnetars could very well have a certain helicity to them.
Could a dipole field with such asymmetries embedded inside the NS have any effect on the aftermath of a BHNS merger? To address this question, in this paper we perform long-term full GRMHD simulations of BHNS mergers where an asymmetric magnetic field dipole is embedded in the NS. 

The rest of the paper is organized as follows.
In Sec.~\ref{sec:level2}, we describe the system and grid setup used in our simulations. We then present the results in Sec.~\ref{sec:level3} along two major directions, 
i.e., the magnetic field amplification and the large-scale coherence of the magnetic field.  Lastly, we summarize our conclusions in Sec.~\ref{sec:level4}. 

\section{\label{sec:level2}System and grid setup}

\begin{figure*}[ht]
\begin{center}
\subfloat[]{%
\includegraphics[scale=0.4]{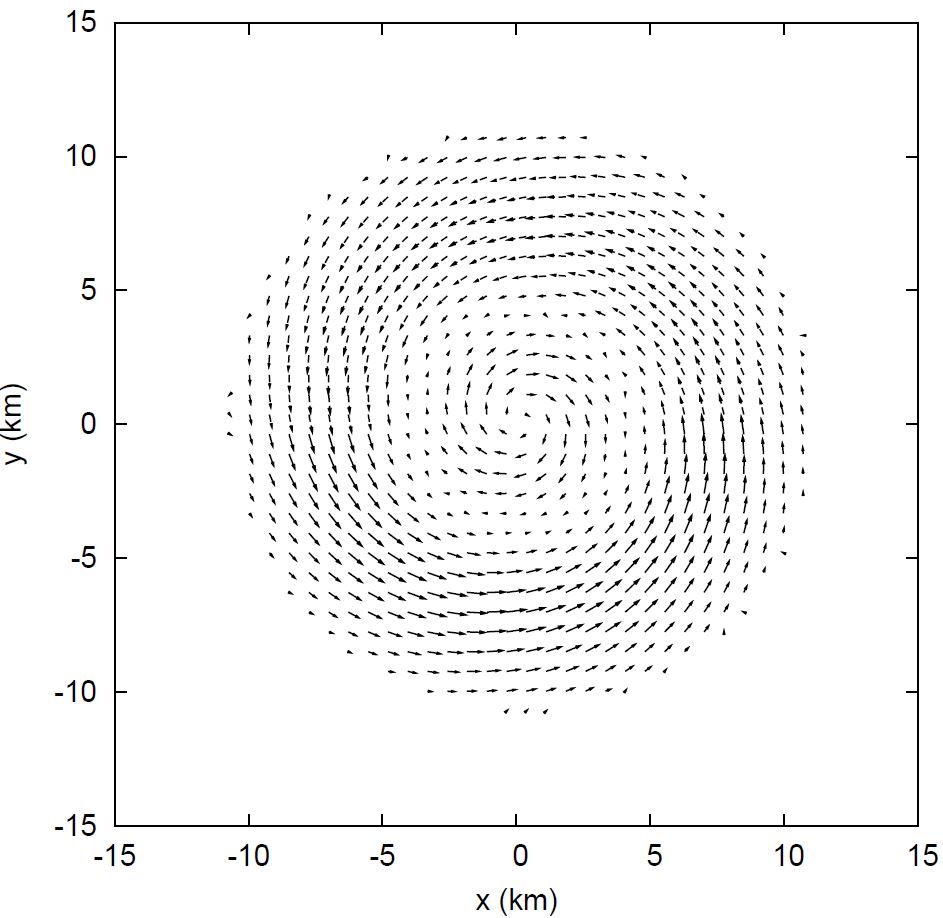}
}%
\subfloat[]{%
\includegraphics[scale=0.32]{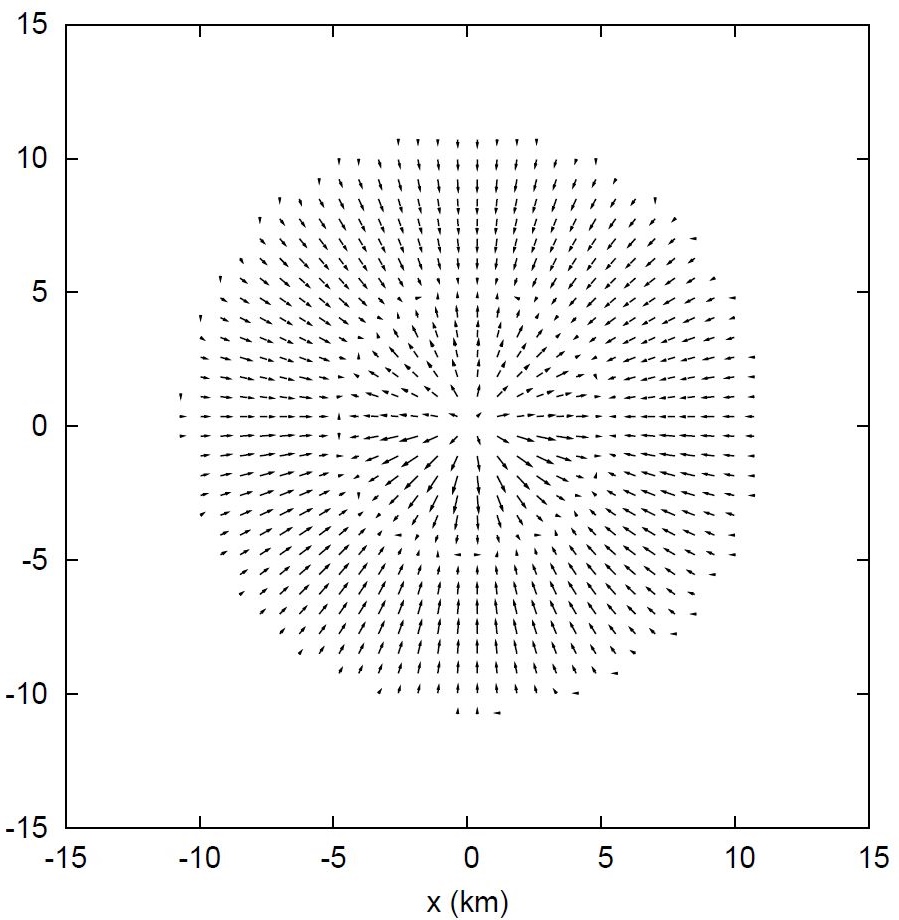}
}%
\end{center}
\caption{
Comparison of the initial magnetic field configuration on the equatorial plane between a) the {\bf Asym60} case and b) the {\bf Sym} case. 
}
\label{fig:magconfigxy}
\end{figure*}

\begin{figure*}[htb]
\begin{center}
\subfloat[]{%
\includegraphics[scale=0.4]{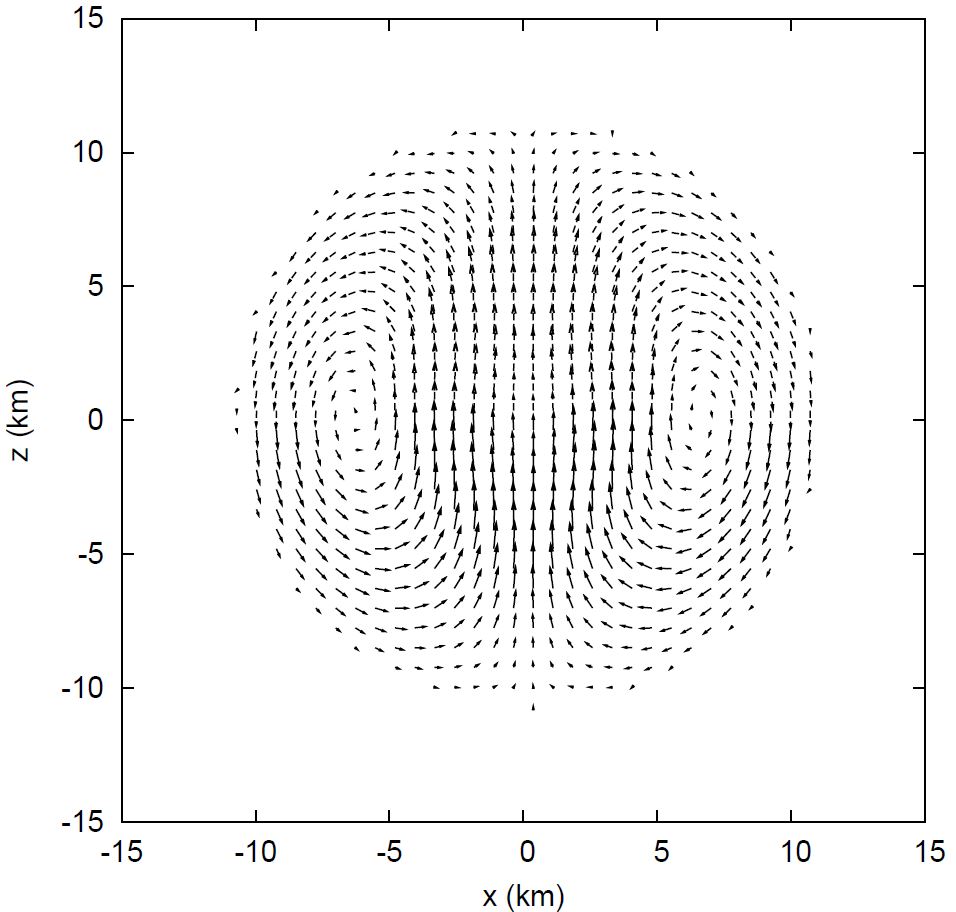}
}%
\subfloat[]{%
\includegraphics[scale=0.32]{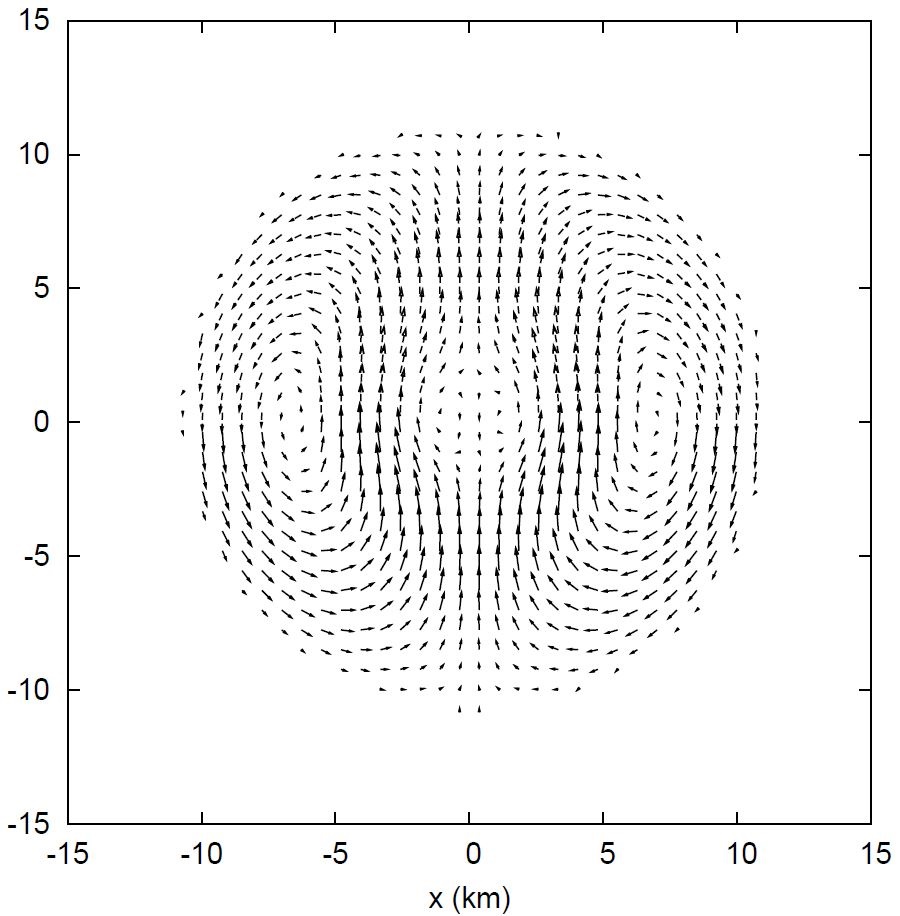}
}%
\end{center}
\caption{
Comparison of the initial magnetic field configuration on the meridional plane between a) the {\bf Asym60} case and b) the {\bf Sym} case.
}
\label{fig:magconfigxz}
\end{figure*}

We simulate BHNS binary mergers with a mass ratio 4, where the BH is spinning at an angular velocity of 0.75 of the extreme Kerr limit. 
The initial data are set up using the quasiequilibrium method as detailed in Ref.~\cite{Kyutoku09}, where a dimensionless orbital angular velocity $m_0\Omega$, is employed to characterize the initial data in lieu of the orbital separation. Here, $m_0=M_{BH}+M_{NS}$, where $M_{BH}$ and $M_{NS}$ are the gravitational masses of the BH and NS respectively.
The BHNS merger in the current study has $m_0\Omega=0.056$.
The NS is modeled with the Akmal-Pandhalipande-Ravenhall equation of state (EOS) \cite{Akmal98} and has a mass of $1.35 M_{\odot}$. The EOS is implemented using the piecewise-polytropic method \cite{Read09} which enables the conversion of plasma kinetic energy into thermal energy via a thermal component of the specific internal energy \cite{Shibata08}. We choose $\Gamma=1.8$ for the thermal part of the gamma-law EOS.

Inside the NS, the magnetic field is seeded in the NS via a vector potential of the form:
\begin{eqnarray}
\label{eq:vecpot}
A_j=&&[-(y-y_{\text{NS}})\sin\theta\cdot\delta^x_j+(x-x_{\text{NS}})\delta^y_j \nonumber\\
    &&+\cos\theta\cdot\delta^z_j]A_{\text{b}} \text{max}(P-P_{\text{c}},0)^2,
\end{eqnarray}
where ($x_{\text{NS}},y_{\text{NS}}$) are the coordinates of the NS center, $P$ is the pressure, $P_{\text{c}}=P(\rho=0.04\rho_{\text{max}})$, where $\rho_{\text{max}}$ is the maximum rest-mass density, and $j=x,y,z$. $A_{\text{b}}$ is set to be $10^{15}$G. The $z$ component in Eq.~(\ref{eq:vecpot}) introduces a helicity to the resulting magnetic field. Decreasing $\theta$ increases the helicity of the dipole, as well as the magnitude of the maximum magnetic field strength by an order of $\sim\sin^{-1}\theta$. The increase in the maximum field strength, plays a negligible role in influencing the dynamics of the system.
We set $\theta=60^{\circ},75^{\circ},90^{\circ}$ and denote them as {\bf Asym60}, {\bf Asym75} and {\bf Sym} respectively. $\theta=90^{\circ}$ gives the symmetric dipole with $A_{\text{b}}$ as the maximum magnetic field strength. In Figs.~\ref{fig:magconfigxy} and ~\ref{fig:magconfigxz}, we show the initial magnetic field configurations on the equatorial and meridional planes respectively, comparing the {\bf Asym60} and {\bf Sym} cases. 

The simulations are performed using the Kyoto group GRMHD code which implements the ideal approximation for the magnetic field \cite{Shibata05,Shibata06,Kiuchi12}. 
In the ideal approximation, the electric field vanishes in the fluid frame moving with 4-velocity $u^{\nu}$, which renders the magnetic field frozen into the fluid. 
The Einstein equations are evolved using the BSSN formalism \cite{Shibata95,Baumgarte98},
with the BH evolution handled using the puncture method \cite{Baker06,Campanelli06}.
As our computational domain, we employ a 3D grid without imposing symmetries. The computational domain is spanned by $453^3$ grid points using eight levels of grid refinement (the grid resolution is refined by a factor of 2 as we go from level 1 to 8). The finest grid patch covers both the BH and NS. The outer boundary of the computational domain is located at $\approx 7700$km. With the finest grid resolution of 0.27km, the NS is covered by 60 grid points across the NS. Further details of the grid structure can be found in Ref.~\cite{Kiuchi12}.

As the binary inspirals, the NS is tidally disrupted by the BH, as expected from previous studies described in Refs.~\cite{Shibata11,Kyutoku13} on BHNS systems with the mass ratio, spin and EOS that we set for this paper. 
The system merges after about three orbits at $\approx 8$ms after the beginning of the simulation. 
From here onwards, we denote the time of merger as $t_{\text{merge}}$.
A massive accretion disk is formed after the merger \cite{Shibata11,Kyutoku13}.
We simulate our system up to $(t-t_{\text{merge}})\approx 100$ms. Although we do not take into account neutrino transport, we estimate that the effects of neutrino viscosity and drag will not severely impact the magnetic field amplification in our system. We show this estimation in Sec.~\ref{sec:level3}. 

\begin{figure*}[t]
\begin{center}
\subfloat[]{%
\includegraphics[width=0.5\textwidth,keepaspectratio=true]{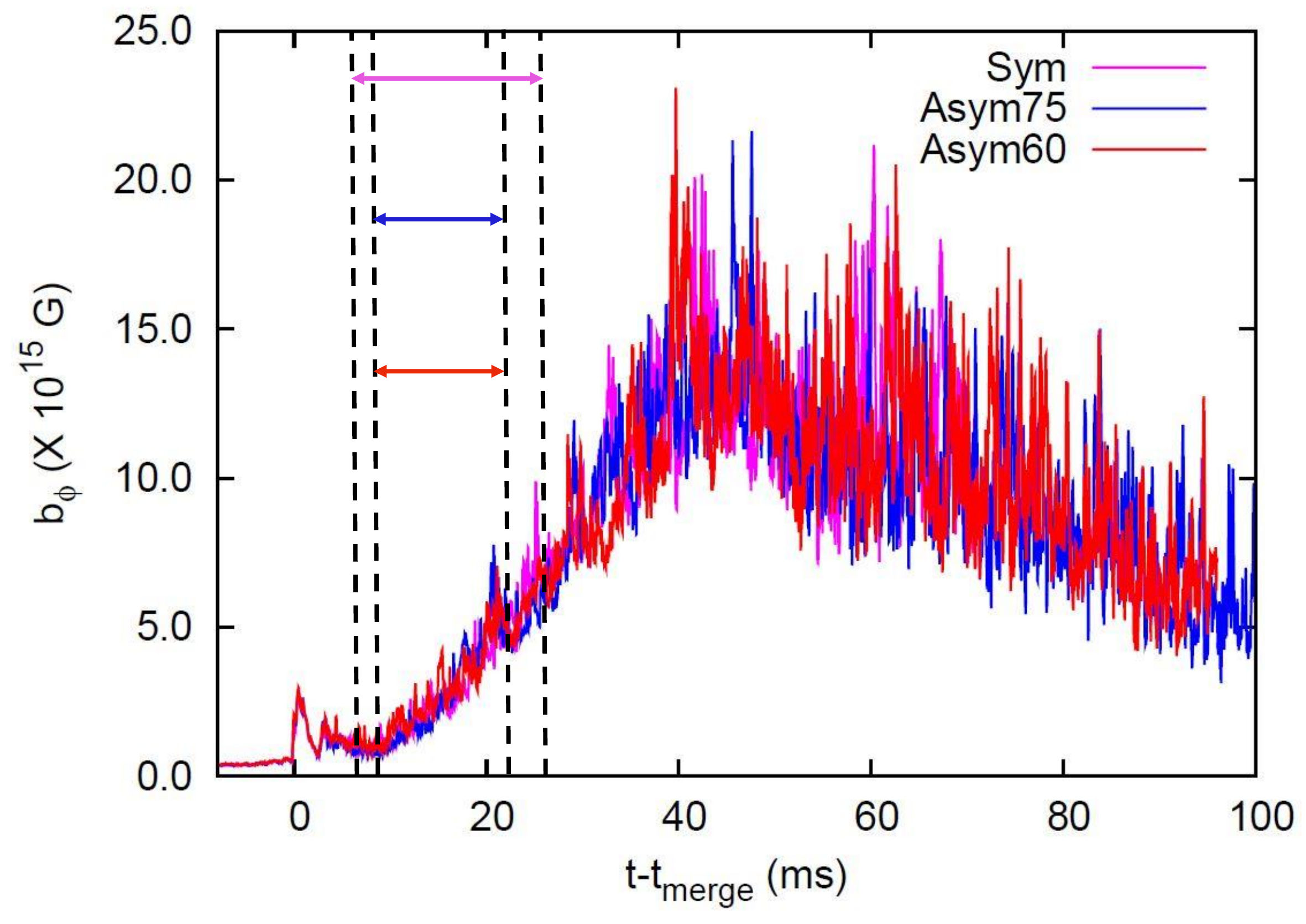}
}%
\subfloat[]{
\includegraphics[width=0.5\textwidth,keepaspectratio=true]{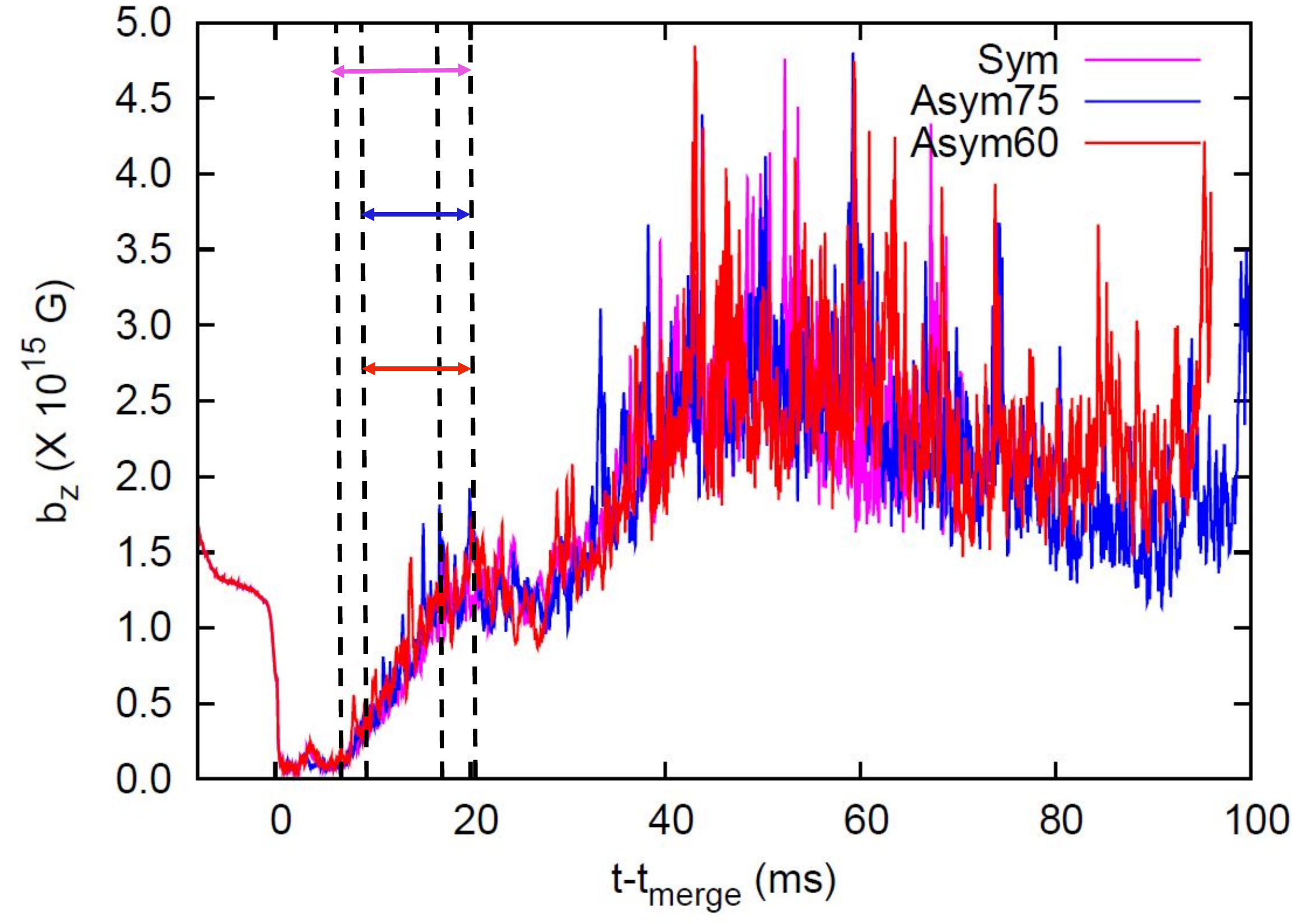}
}
\end{center}
\caption{
The growth of a) the toroidal field and b) the vertical field, both measured in the fluid frame, for all cases.
The dashed lines demarcate the range where the growth is fitted with an exponential function of the form $\exp{(a(t-t_{\text{MRI}}))}$,
where $a$ is the growth rate, and $t_{\text{MRI}}$ is the time when the MRI is triggered.
The pink arrow labels the range for the {\bf Sym} case, blue for {\bf Asym60} and red for {\bf Asym75}. 
}
\label{fig:compb}
\end{figure*}

\begin{figure*}[ht]
\centering
\subfloat{\includegraphics[angle=-90,scale=1.0]{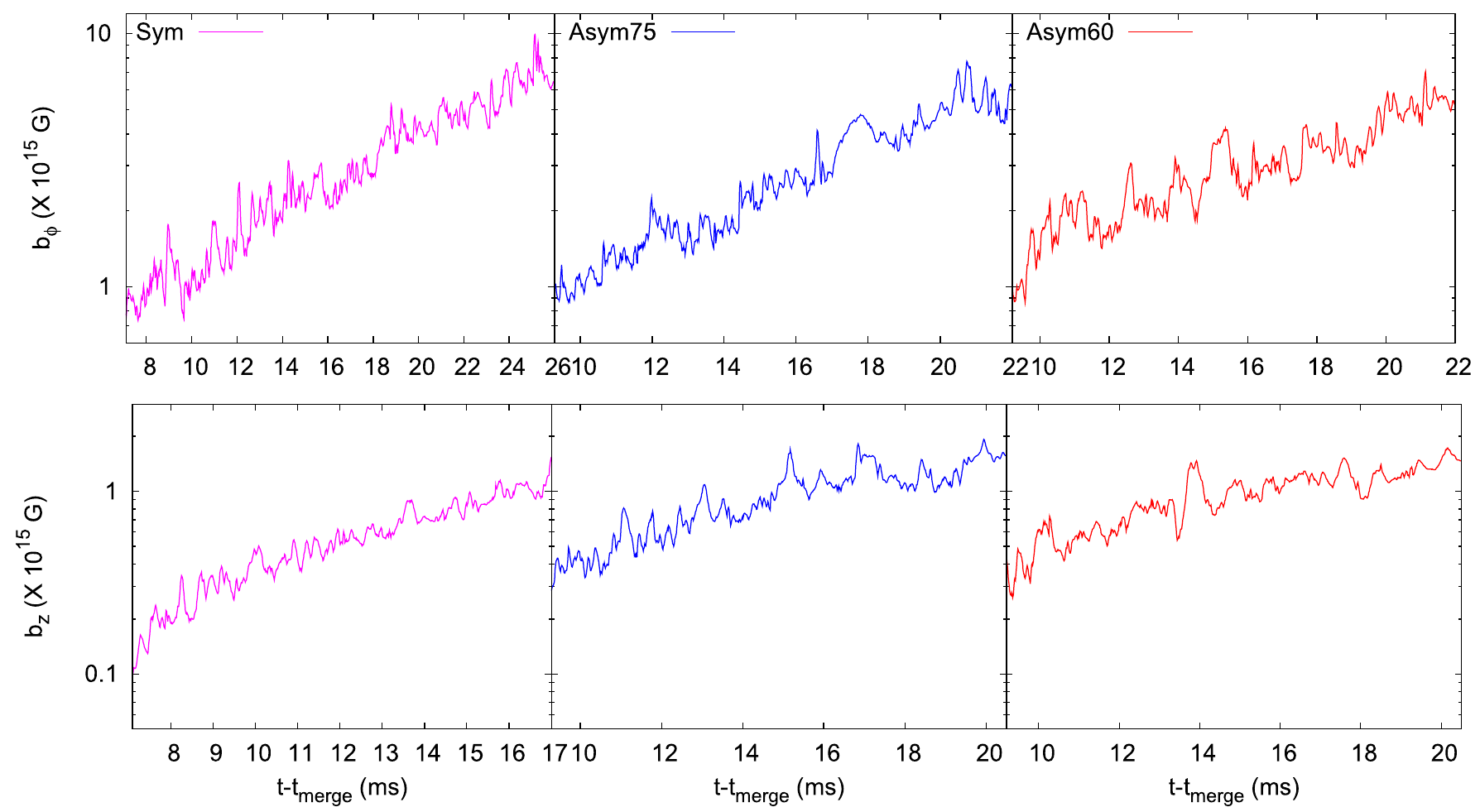}}
\caption{
The close-up of the exponential growth phase of the toroidal field (top panel) and the vertical field (bottom panel), both measured in the fluid frame, for all cases.
}
\label{fig:compbphibzclose}
\end{figure*}

\begin{figure*}[ht]
\begin{center}
\subfloat[]{%
\includegraphics[scale=0.5]{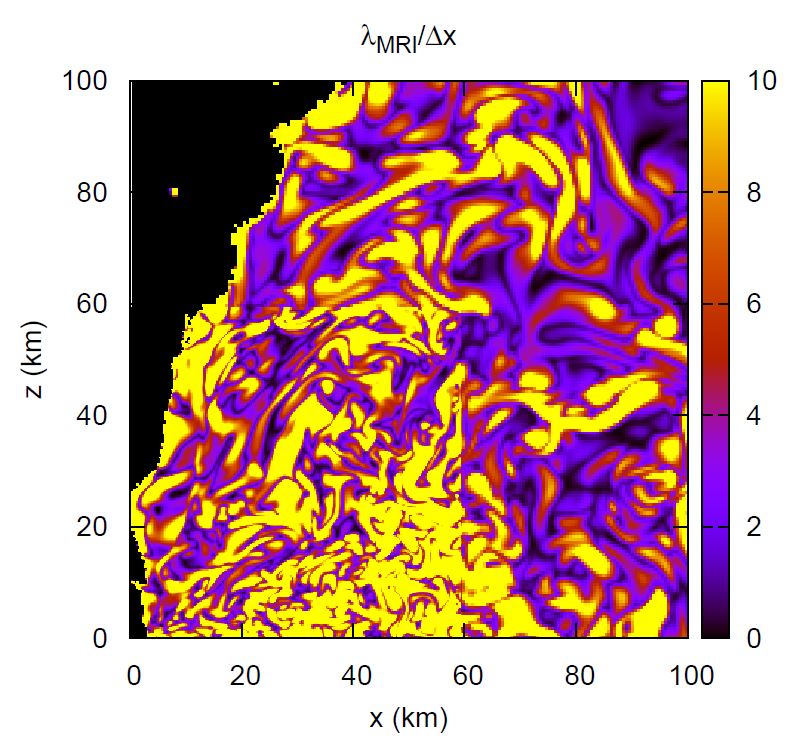}
}%
\subfloat[]{
\includegraphics[scale=0.5]{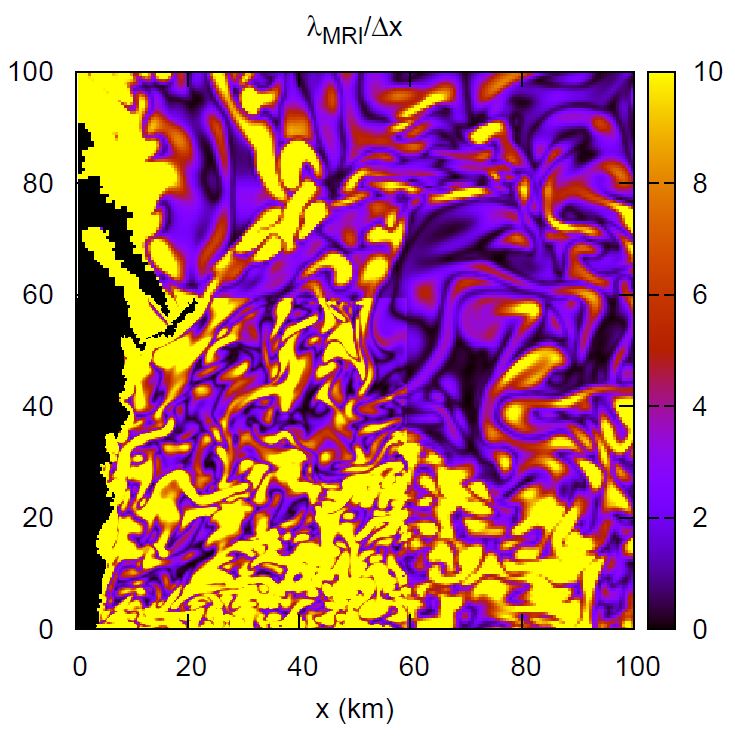}
}
\end{center}
\caption{
Snapshots of $Q_{\phi}$ (cf. Eq.~\ref{eq:resoeq}) on the meridional plane for a) {\bf Sym} and, b) {\bf Asym60} at $ t-t_\text{merge}\approx 18$ms.
}
\label{fig:resobphi18}
\end{figure*}

\begin{figure*}[ht]
\begin{center}
\subfloat[]{%
\includegraphics[scale=0.5]{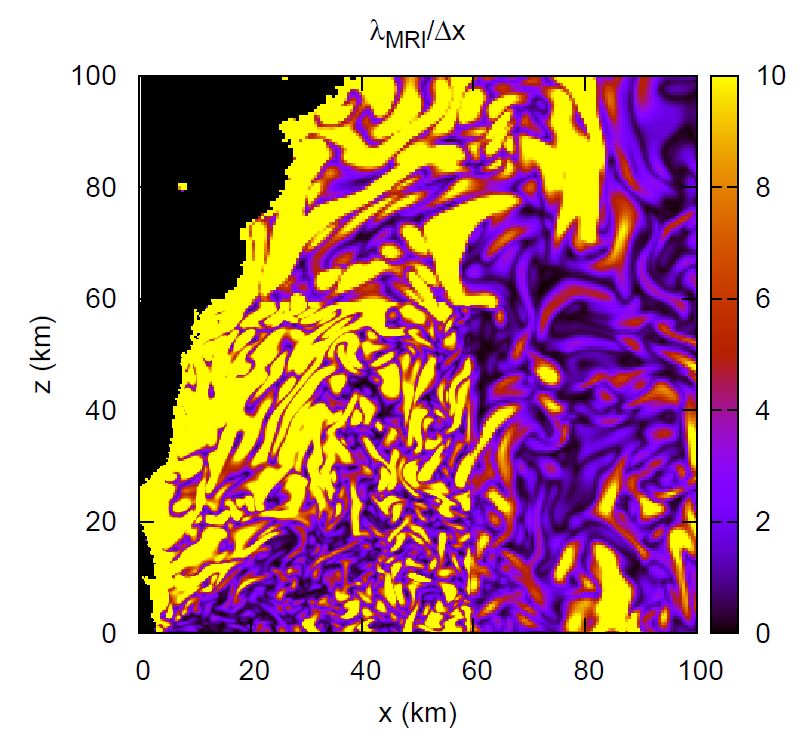}
}%
\subfloat[]{
\includegraphics[scale=0.5]{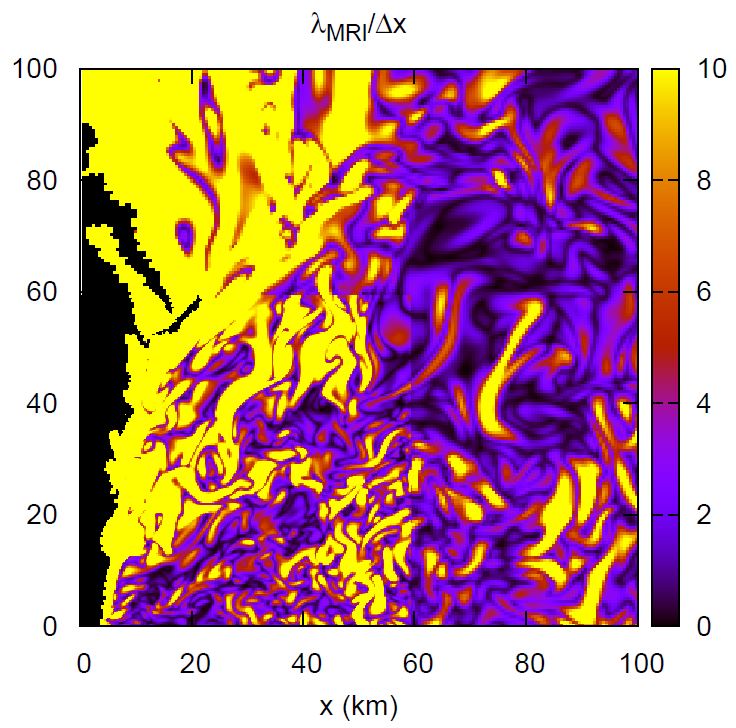}
}
\end{center}
\caption{
Snapshots of $Q_z$ (cf. Eq.~\ref{eq:resoeq}) on the meridional plane for a) {\bf Sym} and, b) {\bf Asym60} at $ t-t_\text{merge}\approx 18$ms.
}
\label{fig:resobz18}
\end{figure*}

\begin{figure*}[ht]
\begin{center}
\subfloat[]{%
\includegraphics[scale=0.5]{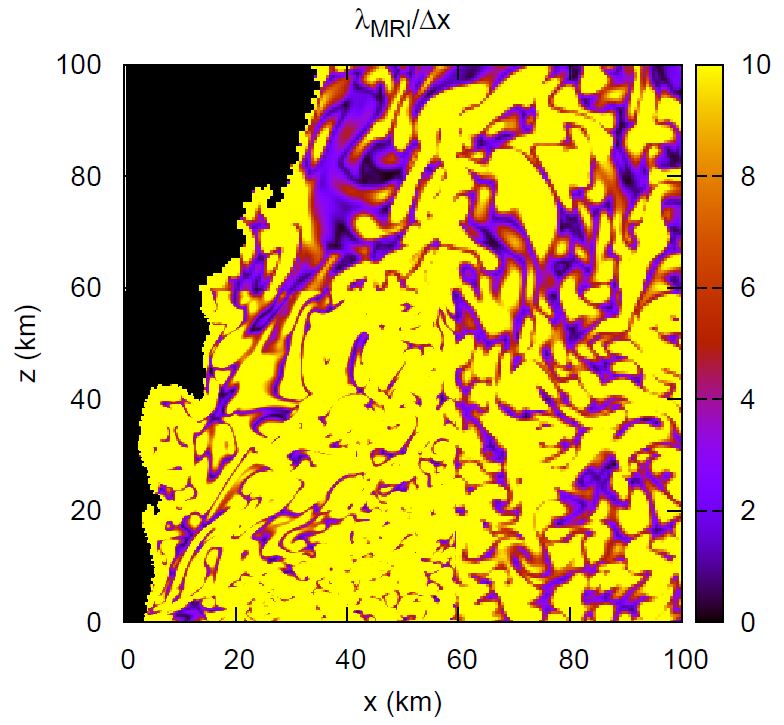}
}%
\subfloat[]{
\includegraphics[scale=0.5]{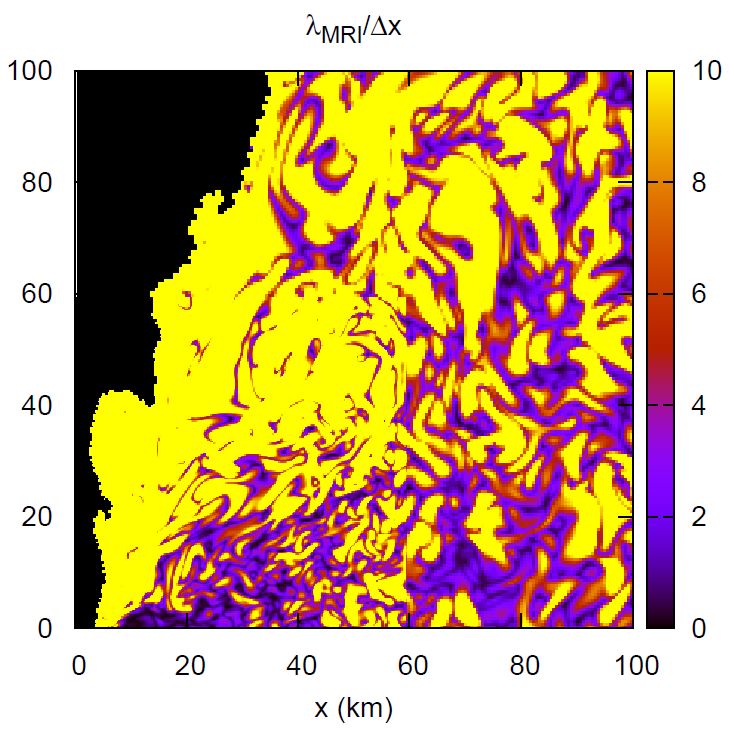}
}
\end{center}
\caption{
Snapshots of a) $Q_{\phi}$ and b) $Q_z$ (cf. Eq.~\ref{eq:resoeq}) on the meridional plane for {\bf Asym60} at $ t-t_\text{merge}\approx 23.5$ms.
}
\label{fig:resob23p5}
\end{figure*}

\begin{figure*}[ht]
\begin{center}
\subfloat[]{%
\includegraphics[scale=1.0]{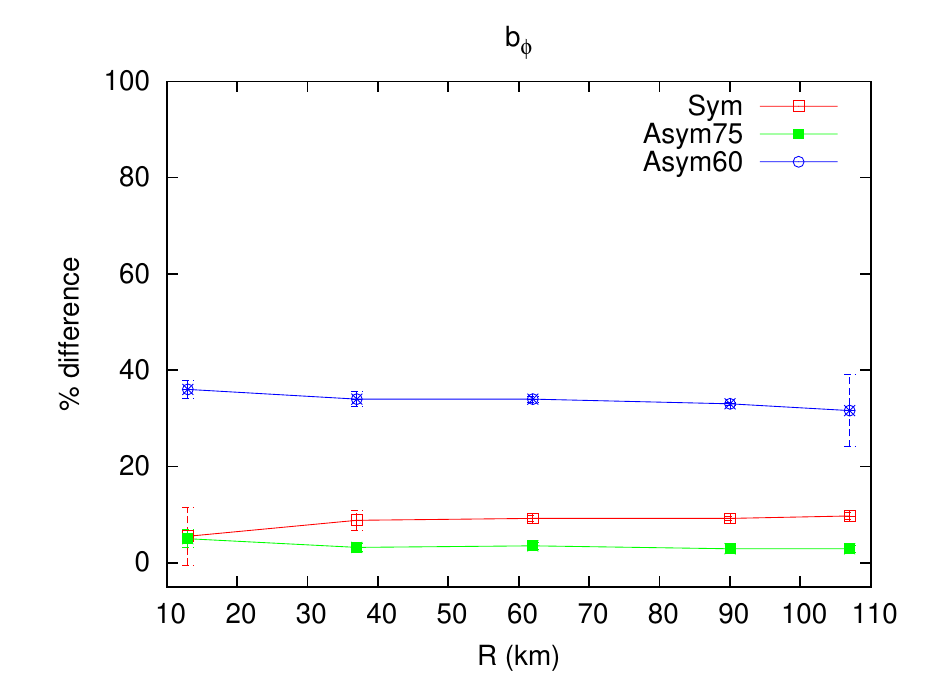}
}
\subfloat[]{
\includegraphics[scale=1.0]{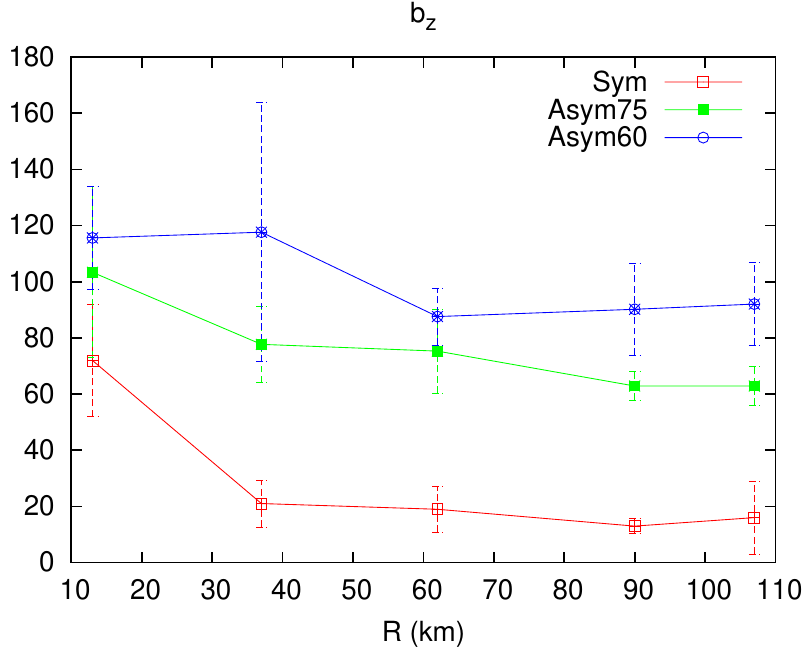}
}
\end{center}
\caption{
Percent differences between growth rates obtained from simulation, with predictions by linear theory of the growth rates of the toroidal and vertical fields on the equatorial plane.
The percent differences shown are averages obtained from points at a radius, $R$, from the center of the system,
whereas the error bars are standard deviations from the average.
}
\label{fig:mribphiz}
\end{figure*}

\begin{figure*}[ht]
\begin{center}
\subfloat[]{%
\label{fig:MRIneu}
\includegraphics[scale=1.0]{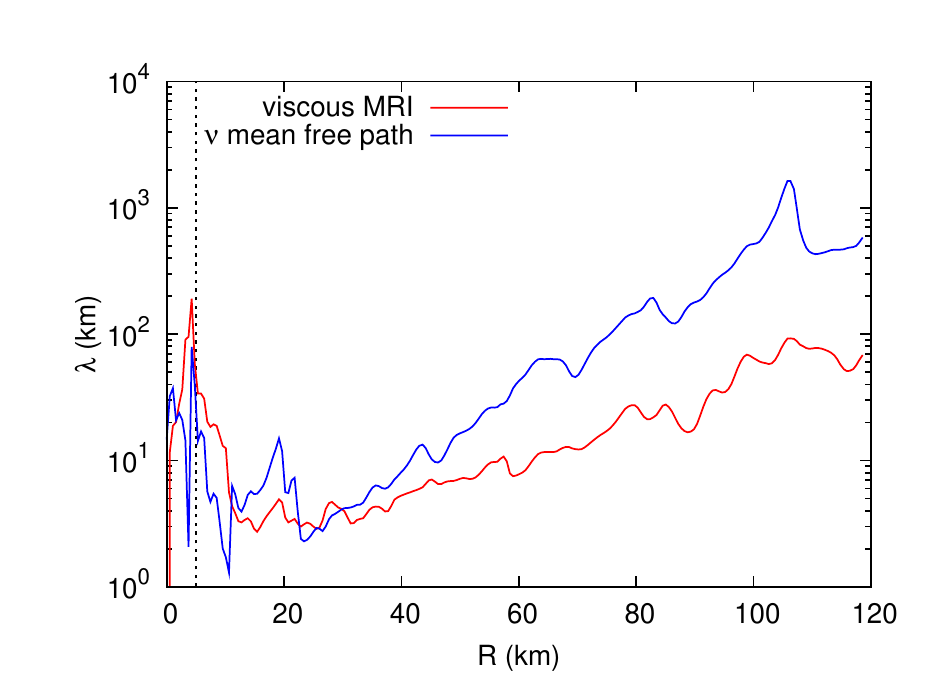}
}%
\subfloat[]{
\label{fig:dampneu}
\includegraphics[scale=1.0]{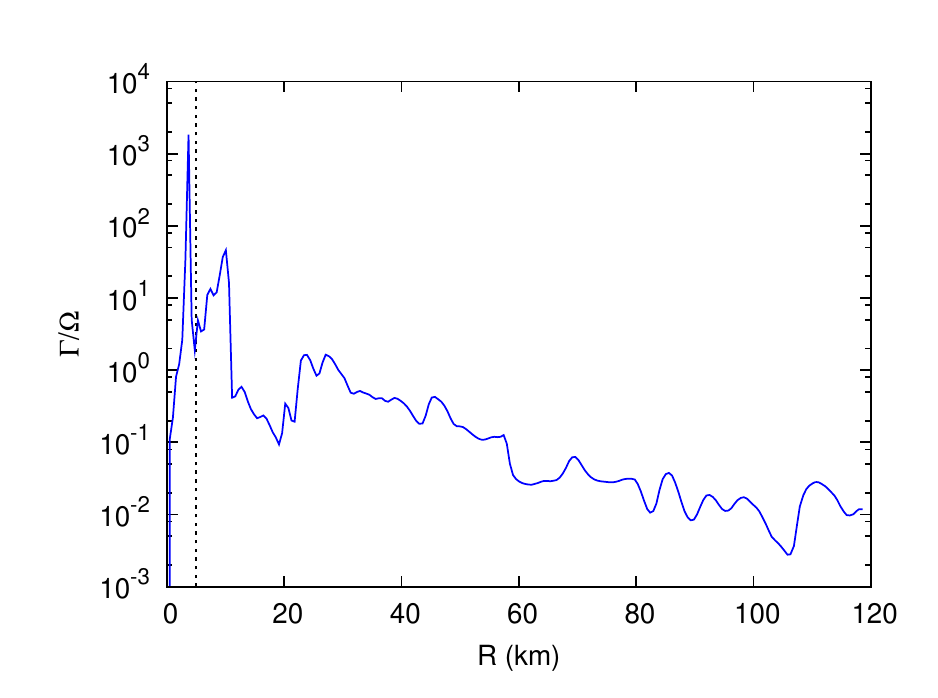}
}
\end{center}
\caption{
a) The comparison between the neutrino mean free path and the viscous MRI wavelength, and b) the damping rate, $\Gamma$, as measured radially throughout the extent of the accretion disk bulk 
for the {\bf Asym60} case.
The dashed lines in the figures indicate the location of the BH apparent horizon. 
}
\end{figure*}

\section{\label{sec:level3}Results}

Up to $t-t_{\text{merge}}\approx 42$ ms, the magnetic field in the accretion disk formed around the BH remnant is amplified by up to a factor of $\approx 20$ in the toroidal component, and $\approx 45$ in the vertical component. As predicted in Ref.~\cite{Balbus92}, the radial component dominates over the vertical component by up to an order of magnitude at saturation. In Fig.~\ref{fig:compb}, we show the evolution of the fluid frame maximum toroidal and vertical fields, respectively. The toroidal field is at first solely amplified as it is wound up by the matter accreting onto the BH during tidal disruption. During this time, the poloidal field which is initially embedded in the NS, decreases sharply as the NS is disrupted at the time of merger. After the formation of the accretion disk, both the toroidal and vertical fields are exponentially amplified by the MRI.
Figure~\ref{fig:compbphibzclose} shows the close-up of the period of exponential amplification, delimited with dashed black lines in Fig.~\ref{fig:compb}, observed in both the toroidal and vertical fields. The order of magnitude of the amplification is similar to that seen in Ref.~\cite{Shibata06a}, with the toroidal field generally an order of magnitude stronger than the vertical field, and the corresponding magnetic field perturbations in the toroidal direction possessing much larger wavelengths than in the vertical direction as expected in a nonaxisymmetric scenario.

In order to clarify whether the MRI is triggered in the cases studied here, we evaluate whether the wavelength of its fastest growing mode is sufficiently resolved in the bulk of the accretion disk.
We follow Refs.~\cite{Etienne12,Kiuchi15} and estimate this wavelength along the $i$th direction as:
\begin{equation}
\label{eq:lambdaMRI}
\lambda_{\text{MRI},i}=|v_{i,A}|/\frac{2\pi}{\Omega} \quad(i=\phi,z)
\end{equation}
where $v_{i,A}$ is the component of the Alfv\'{e}n speed in the $i$th direction in the \textit{fluid frame}, and is given as $v_{i,A}=b_i/\sqrt{4\pi\rho h+b^2}$.
$b_i$ is the magnetic field measured in the fluid frame, $\rho$ is the rest-mass density and $h$ is the specific enthalphy.
$\Omega$ is the angular velocity of the plasma in the accretion disk, calculated as:
\begin{eqnarray}
\label{eq:Omegadef}
\Omega = \frac{u^{\phi}}{u^0} & = & \alpha \tilde{v}^{\phi}-\beta^{\phi} \nonumber\\
                              & = & \alpha \left(\frac{-y \tilde{v}^x+x \tilde{v}^y}{r^2}\right)-\left(\frac{-y \beta^x+x \beta^y}{r^2}\right) \nonumber\\
                              & = & \frac{(-y v^x+x v^y)}{r^2},
\end{eqnarray}
where $u^{\phi}$ and $u^0$ are components of the fluid 4-velocity, $\tilde{v}^i (i=\phi,x,y)$ is the component of the fluid 3-velocity relative to Eulerian observers along the azimuthal, $x$ and $y$ directions, $\beta^i (i=\phi,x,y)$ a component of the shift vector along the azimuthal, $x$ and $y$ directions, $v^i (i=x,y)$ the coordinate velocities of the fluid in the $x$ and $y$ directions, and $r^2=x^2+y^2$, with $x$ and $y$ being the coordinate distances measured from the grid domain center.
All modes with wavelengths above $\lambda_{\text{MRI},i}$ are unstable, with growth rates lower than that for the mode with $\lambda_{\text{MRI},i}$.
We quantify whether MRI is triggered by using the following diagnostic:
\begin{equation}
\label{eq:resoeq}
Q_i=\frac{\lambda_{\text{MRI},i}}{\Delta x} \quad(i=\phi,z).
\end{equation} 
Following the criteria presented in Refs.~\cite{Balbus98,Noble10,Hawley11,Bicknell13}, we designate $Q_i\gtrsim10$ for adequate modeling of turbulence and magnetic field amplification due to the MRI.
We measure $Q_i$ at $t-t_\text{merge}\approx 18$ms which lies within the exponential growth phase for the toroidal field as shown in Figs.~\ref{fig:compb} and~\ref{fig:compbphibzclose} and observe wide regions where $Q_{\phi}$ \textit{significantly exceeds} $10$. Patches of the flow where $Q_{\phi}<10$ exist and these are interspersed with regions where $Q_{\phi}\gg 10$ given that the accretion flow is highly turbulent. Since the time of the measurement is well past the beginning of the exponential growth phase, the wide regions with $Q_{\phi}\gg 10$ indicate that increasingly larger wavelengths of the fastest mode could become unstable as the magnetic field is amplified with time by a combination of magnetic winding and MRI. The MRI that can be triggered is expected to be nonaxisymmetric since the BHNS systems we consider here involve the toroidal motion of accreting matter onto the BH, producing mostly toroidal fields. Though with vertical field components however small, the growth rate of the MRI can be greatly increased \cite{Balbus98,Hawley11}. As we mentioned in Sec.~\ref{sec:level1}, in the limit where the magnetic field perturbation wave number in the vertical direction greatly exceeds the toroidal one, the growth rate approaches the maximum growth rate for the axisymmetric MRI \cite{Balbus98,Hawley11}, which greatly exceeds the growth rate predicted for a purely toroidal field. Thus, an increased $Q_z$ is expected to enable the capture of the maximum growth rate for the nonaxisymmetric system. In the present study, the area with $Q_z\gtrsim10$ is still quite limited to the corona and the disk edges abutting it, well into the late stage of the exponential growth phase, with much of the bulk of the accretion disk still underresolving the fastest growing modes in the vertical direction as seen in Fig.~\ref{fig:resobz18}. Given this consideration, the fastest modes would still be elusive with the grid spacing of $0.27$km on the finest grid, even though the results of our present study indicate that certain qualitative features of the nonaxisymmetric MRI, of note the sustained turbulence generated by sufficiently wide regions with $Q_{\phi}\gg 10$ \cite{Fromang06,Flock11,Hawley11,Bicknell13}, are captured in our simulations. 

In Figs.~\ref{fig:resobphi18} and ~\ref{fig:resobz18}, we compare the meridional profiles of $Q_{\phi}$ and $Q_z$ respectively, between the symmetric case {\bf Sym} and the most asymmetric case considered in this paper {\bf Asym60}. In both cases, the bulk of the accretion disk achieves $Q_{\phi}\gtrsim 10$, whereas the area with $Q_z\gtrsim 10$ is confined to the corona and the disk edges abutting it. We again note that the measurement is performed at a time well past the beginning of the exponential growth phase observed in Figs.~\ref{fig:compb} and~\ref{fig:compbphibzclose}. As such, we do not expect that the wide region with $Q_{\phi}\gtrsim10$ at this time is the sole contributor to the amplification of the magnetic field from the beginning, but rather, it is an indication that increasingly larger wavelengths of the fastest mode could become unstable as the magnetic field is amplified with time. The symmetric case here, at a resolution of $0.27$km on the finest grid, yields consistent results with that presented in Ref.~\cite{Kiuchi15}. In the latter study, the wide region in the bulk of the accretion disk shown with $Q_{\phi}\gtrsim 10$ is seen also much after the beginning of the exponential growth phase (shown in the ratio of magnetic field over internal energies given in Fig.~2(c) in Ref.~\cite{Kiuchi15}).  Looking further into the comparison between the symmetric and asymmetric cases in the current study, we observe that the asymmetric case achieves $Q_z\gtrsim 10$ for a slightly larger area in the bulk of the accretion disk as well as in the corona, particularly in the early stages of mass ejection. At late stages of the mass ejection, no perceived difference could be detected in the MRI resolution area between the symmetric case and the asymmetric cases. In the toroidal direction, no difference in the MRI resolution area seems to be apparent even in the early stages. Figure~\ref{fig:resob23p5} shows the MRI resolution at $t-t_\text{merge}\approx 23.5$ms for the {\bf Asym60} case. The area where the MRI fastest mode wavelengths are resolvable has spread considerably with time as compared with that at $t-t_\text{merge}\approx 18$ms.
At $t-t_\text{merge}\approx 23.5$ms, almost the entire area in the bulk of the accretion disk together with the corona, achieves $Q_{\phi}\gtrsim 10$ and $Q_z\gtrsim 10$ respectively, and this picture is similar to the symmetric case at $t-t_\text{merge}\approx 23.5$ms. As mentioned above, this means that increasingly larger wavelengths of the fastest mode could become unstable as the magnetic field is amplified with time, but the growth of the magnetic field may enter the nonlinear regime and saturation. 

Here, we analyze the growth of the magnetic field locally to compare with predictions by the linear theory of MRI.
Predictions by linear theory are computed following Ref.~\cite{Balbus92}, where two coupled second-order ordinary differential equations (ODEs) derived from the linearized MHD equations, are solved for $B_z$ and $B_R$. $B_\phi$ is obtained from the coupling to $B_z$ and $B_R$ via the divergence-free constraint.
The power law decay of the angular velocity, $\Omega$, with respect to the distance from the axis of rotation of the BHNS, $R$, is fitted with the following function \cite{Balbus92}:
\begin{equation}
\label{eq:powerlaw}
\Omega(R)\sim R^{-p/2},
\end{equation}
where $p$ is obtained from the fitting and then input as a parameter in the ODEs.
We also measured the magnetic field, angular velocities and densities at 40 points across the equatorial plane, which are input as initial data for the ODEs.
The 40 points taken span the radii from 13 km to 107 km from the center of the system, along both the azimuthal and the radial directions. 
To quantify magnetic field perturbations, we first obtain an averaged value of the magnetic field at a point, defined as the local nonzero background magnetic field, by employing a smoothening kernel.
The size of the kernel used in the averaging procedure is given by the number of grid points spanned by the \textit{uncorrected} fluctuation in the field, i.e., the fluctuation from a \textit{zero} local background magnetic field, with the greatest spatial extent along a given direction.
The perturbation at a given point is computed as the difference between the measured and the averaged value at that point.
Fourier transforms of the magnetic field in the region mentioned above are then performed along the azimuthal, radial and vertical directions.
Dominant modes are observed from the Fourier transforms, and the corresponding wave numbers are input as parameters for solving the ODEs as well. In effect, for each given point, we have eight dominant modes constructed from the two dominant wave numbers in the azimuthal, radial, and vertical directions respectively.
The ODEs are then integrated using a fourth-order Runge-Kutta method. As a verification, we ensure that our integration scheme reproduces the results presented in Ref.~\cite{Balbus92}. 

The growth rates of $B_\phi$ and $B_z$ resulting from the linear analyses of these dominant modes are approximated using an exponential function as follows:
\begin{equation}
\label{eq:omega}
B_i=B_{i,0}\exp(\omega t) \quad(i=\phi,z),
\end{equation}
where $B_{i,0}(i=\phi,z)$, is the magnetic field at the time when MRI amplification begins, $\omega$ is the approximated growth rate and $t$ is the time measured as the number of orbits of the fluid flow beginning from the time when the magnetic field starts to grow. For linear models which generate multiple growth phases, we consider only the first growth phase.
These growth rates are then compared with the approximated growth rates obtained from the numerical simulations.
The approximation for the numerical simulations are performed over the time ranges as indicated in Figs.~\ref{fig:compb} and ~\ref{fig:compbphibzclose}, i.e., the periods where we observe exponential growth in the field.
To narrow down the modes that most likely produced the growth rates seen in the simulations,
for each of the 40 points considered above, we consider two analytically predicted growth rates that are closest to those observed in the simulations. 
We note that the modes represented by these analytically predicted growth rates are the dominant ones obtained from the Fourier transforms mentioned above. 
Percent differences are then calculated and averaged for a particular distance from the center of the BH-torus system. 
We plot in Fig.~\ref{fig:mribphiz} the percent differences with respect to distance, 
between the growth rates of the maximum toroidal and vertical fields obtained from the simulations, and those predicted by linear theory.
Combined with error bars calculated as the standard deviations from the mean percentage difference obtained from the sample,
the figure is intended to reflect how closely the amplification of the magnetic field could be described by the linear theory of the MRI.

From these figures, we see that the growth of the toroidal field for the {\bf Sym} and {\bf Asym75} cases agrees quite well with that predicted by the linear nonaxisymmetric MRI analyses, with a discrepancy of $\lesssim 10\%$ consistent throughout the equatorial plane of the disk. The growth rates for the most helical case, {\bf Asym60}, deviate from linear theory predictions at $\sim 35\%$ consistent throughout the equatorial plane of the disk, which could be attributed to the high asymmetricity of the magnetic field.  
For the vertical field, the growth rates deviate strongly from linear theory, even extending to $>100\%$, with linear theory mostly overestimating the growth rate.
The deviation increases as the asymmetry in the initial magnetic field topology increases.
This is caused by the poor resolvability of the vertical field on the equatorial plane, as mentioned above.
However, we note in addition that the growth of the magnetic field could also enter the nonlinear regime, which either enhances or hampers growth leading to deviations from linear theory.
This is certainly true in all the cases studied here, where the turbulent state is realized.

Both the toroidal and vertical fields achieve saturation beginning at $t-t_\text{merge}\approx 42$ms.
The saturation levels for all the cases are approximately similar,
with the saturated state for the toroidal field decaying faster than that for the vertical field.
The global magnetic energy density of the vertical field at saturation is $0.1\%$-$0.2\%$ of the kinetic energy of plasma, while that of the toroidal field is $0.4\%$-$0.6\%$.

Over the simulated time scale, neutrino cooling is expected to occur, and indeed, would be the main mechanism whereby efficient accretion could take place to produce the observed luminosities for sGRBs \cite{Nakar07}. However, neutrino radiation would also potentially hamper MRI growth via viscosity and drag \cite{Guilet15}. Following the estimates given in Ref.~\cite{Guilet15} for the neutrino viscosity, the neutrino mean free path, and the wavelength of the fastest growing MRI mode in the viscous regime, here we gauge the effect of neutrino radiation on the MRI growth seen in our study. We follow Refs.~\cite{Shibata08,Kiuchi15} in estimating the temperature of the plasma from the thermal component of the specific internal energy, where the plasma is assumed to be consisting of gas, photons, and relativistic electrons and positrons. Figure~\ref{fig:MRIneu} shows the radial profile of the neutrino mean free path and the viscous MRI wavelength measured along the $x$ axis spanning the bulk of the accretion disk in the {\bf Asym60} model at $t-t_\text{merge}\approx 12$ms, during the MRI growth phase. 
We see that throughout almost the entirety of this range, with the exception of a small region very near the BH, the former exceeds the latter, indicating that the MRI growth would not be affected by neutrino viscosity. Rather, the MRI growth would be in the neutrino drag regime if the damping rate of the drag force is less than the angular velocity of the accretion flow \cite{Guilet15}. Since the damping rate in our case is shown to be orders of magnitude less than the angular velocity, again with the exception of a small region near the BH [Fig.~\ref{fig:dampneu}], we gauge that the MRI growth seen in our study would be put squarely in the ideal MHD regime without being hampered significantly by neutrino radiation as described in Ref.~\cite{Guilet15}.

\begin{figure}[ht]
\begin{center}\includegraphics[scale=1.0]{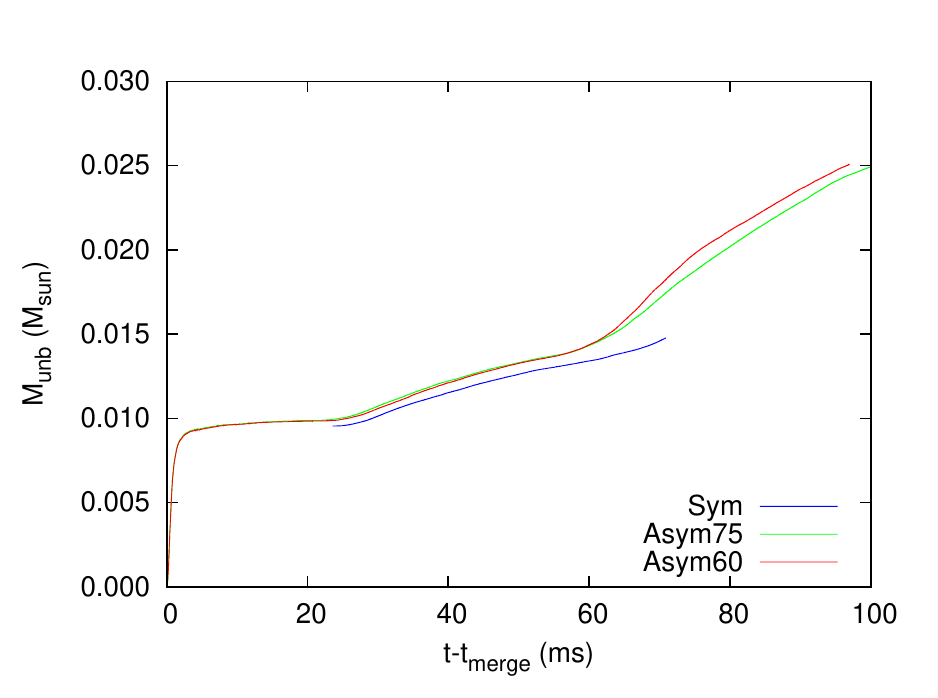}
\caption{Evolution of the unbound mass, $M_{\text{unb}}$ for all cases. For the unbound mass at $t-t_\text{merge}\lesssim25$ms for {\bf Sym}, we refer the reader to Fig.~2(a) of Ref.~\cite{Kiuchi15}, where the system with a resolution of $\Delta x=270$m is exactly the same as the {\bf Sym} case in the current paper.}
\label{fig:unbound}
\end{center}
\end{figure}

\begin{figure*}[ht]
     \begin{center}
        \subfloat[]{%
            \label{fig:wind}
            \includegraphics[scale=0.43]{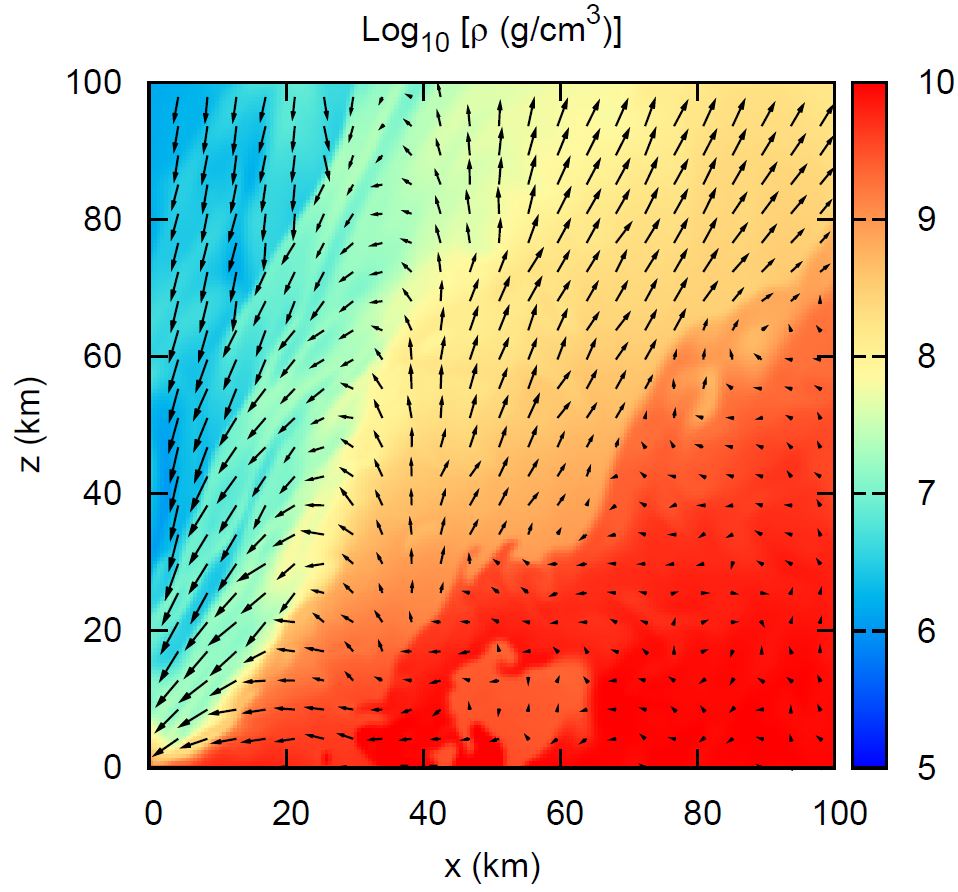}
        }%
        \subfloat[]{
           \label{fig:plasmabet}
           \includegraphics[scale=0.43]{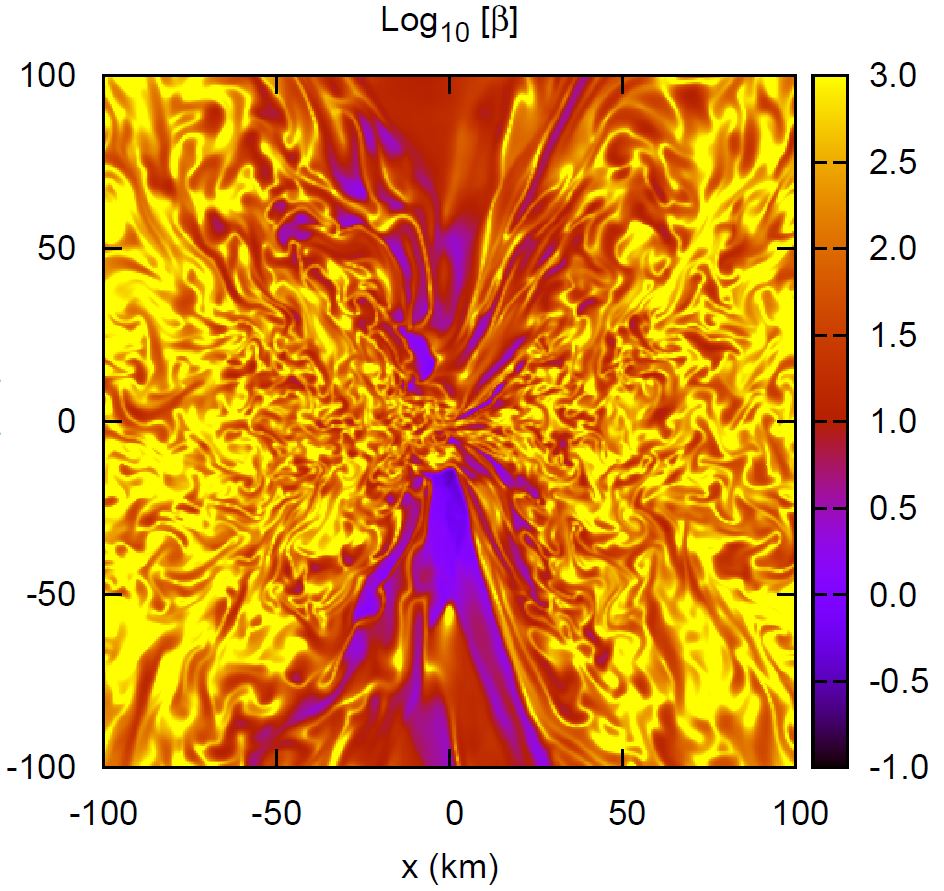}
        }  
    \end{center}
    \caption{
        Profiles of a) rest-mass density and velocity, and b) the plasma $\beta$, on the meridional plane, for {\bf Asym60} at $t-t_\text{merge}\approx 72$ms.
     }
\end{figure*}

\begin{figure}[ht]
\begin{center}\includegraphics[scale=1.0]{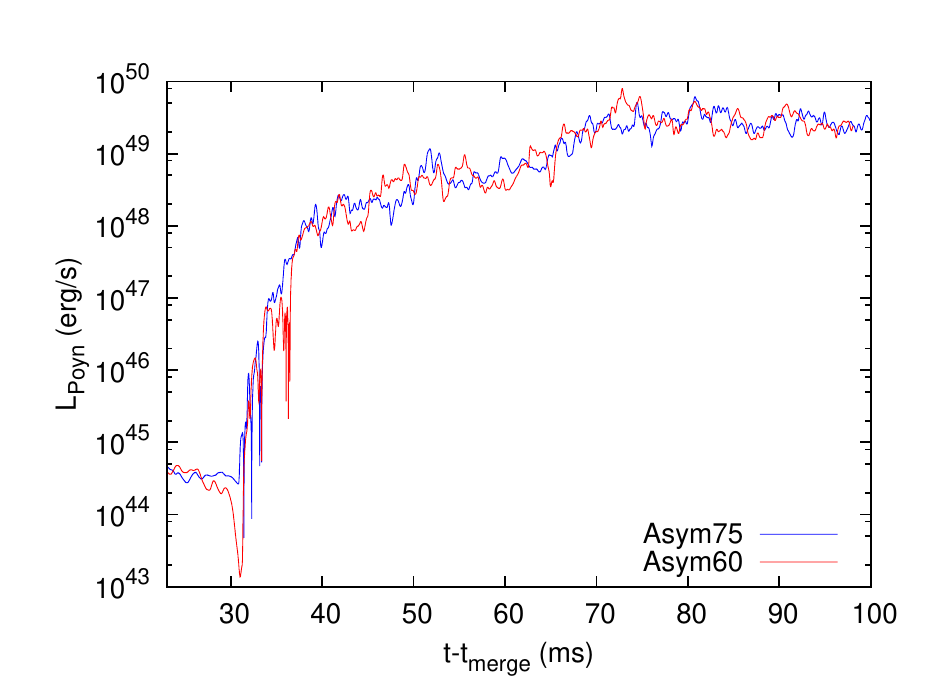}
\caption{The Poynting flux measured at $\approx 960$km from the center of the system, as a function of time for {\bf Asym60} and {\bf Asym75}.}
\label{fig:poynflux}
\end{center}
\end{figure}
\begin{figure}[ht]
\begin{center}\includegraphics[scale=0.43]{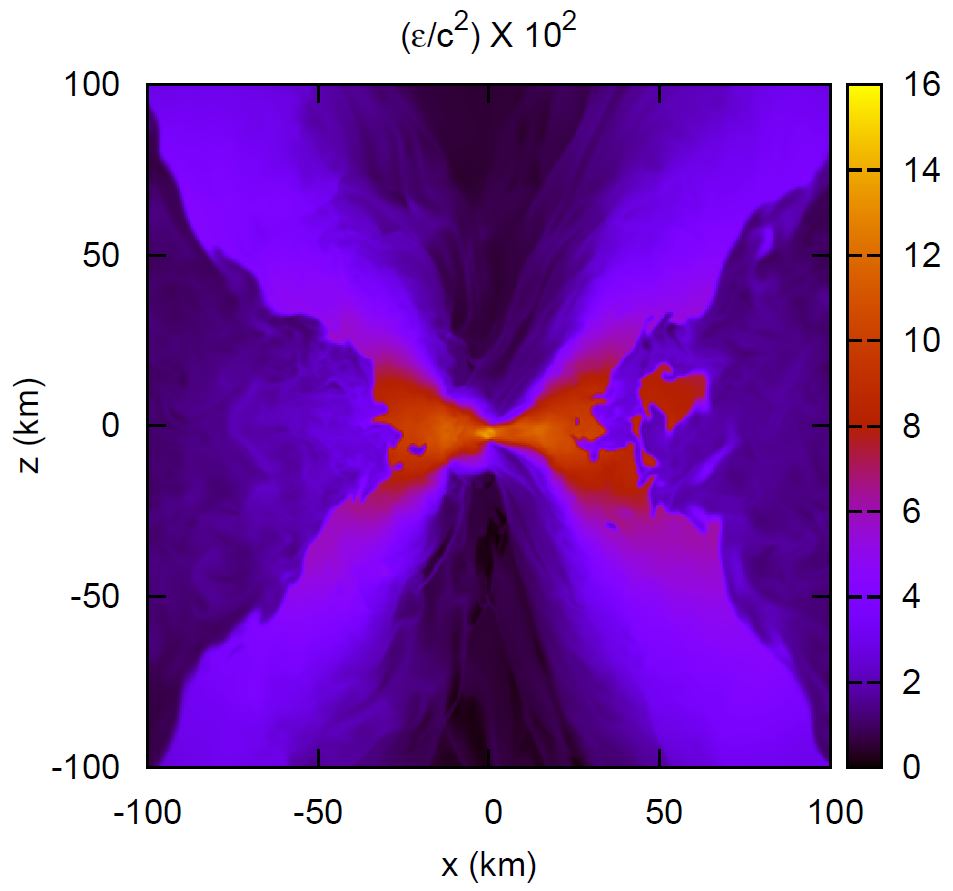}
\caption{
Profile of the thermal component of the specific internal energy, $\epsilon_{th}$, on the meridional plane, for {\bf Asym60} at $t-t_\text{merge}\approx 72$ms.
}
\label{fig:therm}
\end{center}
\end{figure}
\begin{figure}[ht]
\begin{center}\includegraphics[scale=1.0]{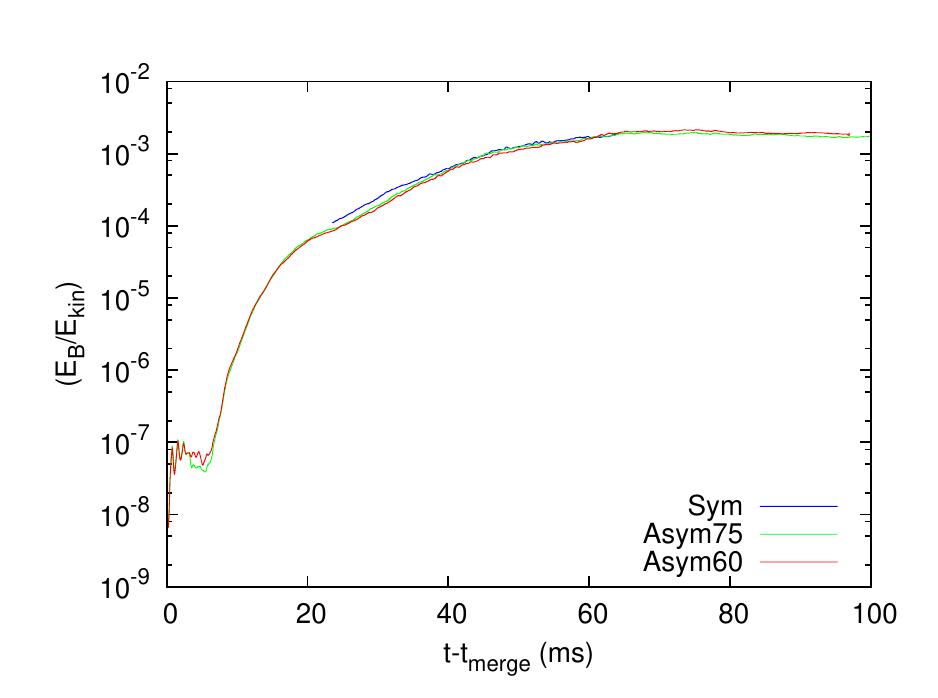}
\caption{The ratio of the global vertical magnetic energy over the plasma kinetic energy measured as a function of time for all cases. At $t-t_\text{merge}\lesssim 25$ms for {\bf Sym}, we refer the reader to the $\Delta x=270$m case in Fig.~2(d) of Ref.~\cite{Kiuchi15} which is exactly the same system as {\bf Sym}. Although the latter shows the ratio of the global magnetic energy over the plasma internal energy instead of over the kinetic energy, the exponential increase at $t-t_\text{merge}\lesssim 25$ms is expected to be similar across these two quantities.}
\label{fig:EBEkin}
\end{center}
\end{figure}

\begin{figure*}[ht]
\centering
\subfloat{\includegraphics[scale=0.8]{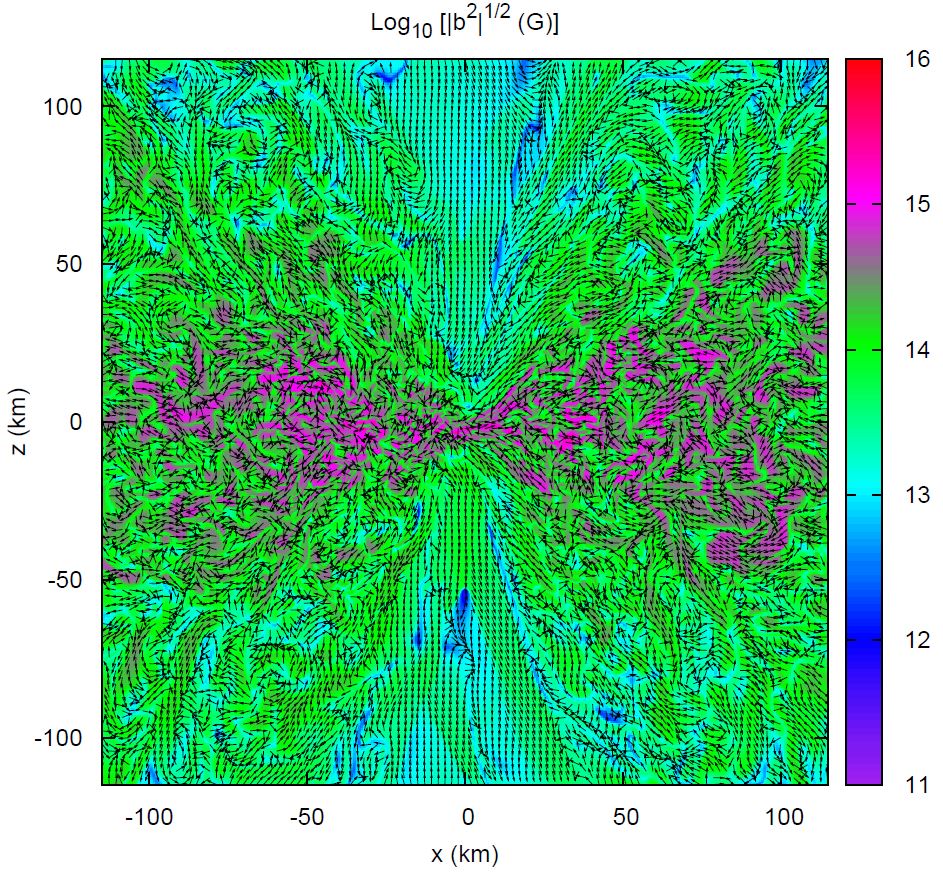}}
\caption{
Close-up of the magnetic vector field on the meridional plane in the disk region around the BH, for {\bf Asym60} at $t-t_\text{merge}\approx 72$ms. Color coding labels the magnetic field intensity, given as $\sqrt{|b^2|}$.
}
\label{fig:coherentclose}
\end{figure*}

\begin{figure}[ht]
\begin{center}\includegraphics[scale=1.0]{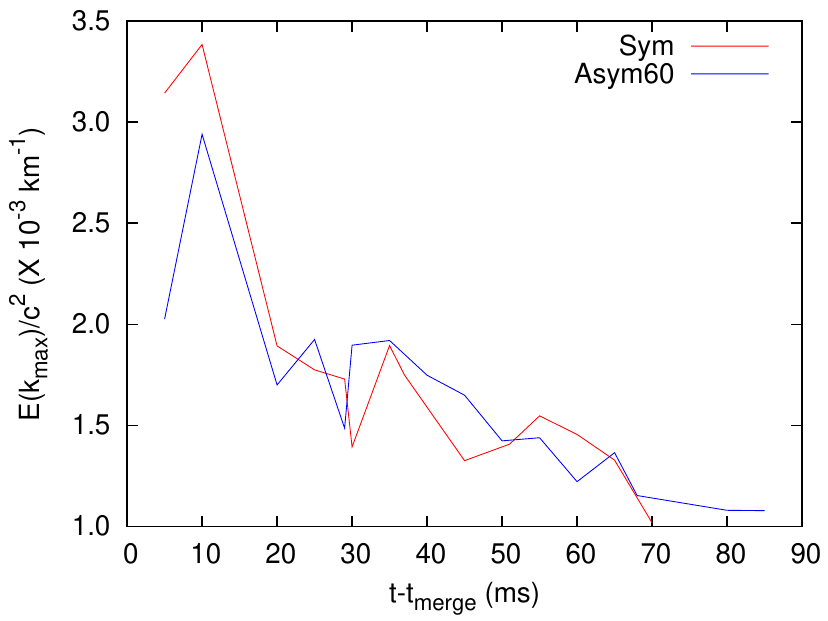}
\caption{
\textit{Mean} amplitude of the dominant mode in the kinetic energy spectrum measured as a function of time for the {\bf Sym} and {\bf Asym60} cases.
}
\label{fig:kmaxt}
\end{center}
\end{figure}
\begin{figure}[ht]
\begin{center}\includegraphics[scale=1.0]{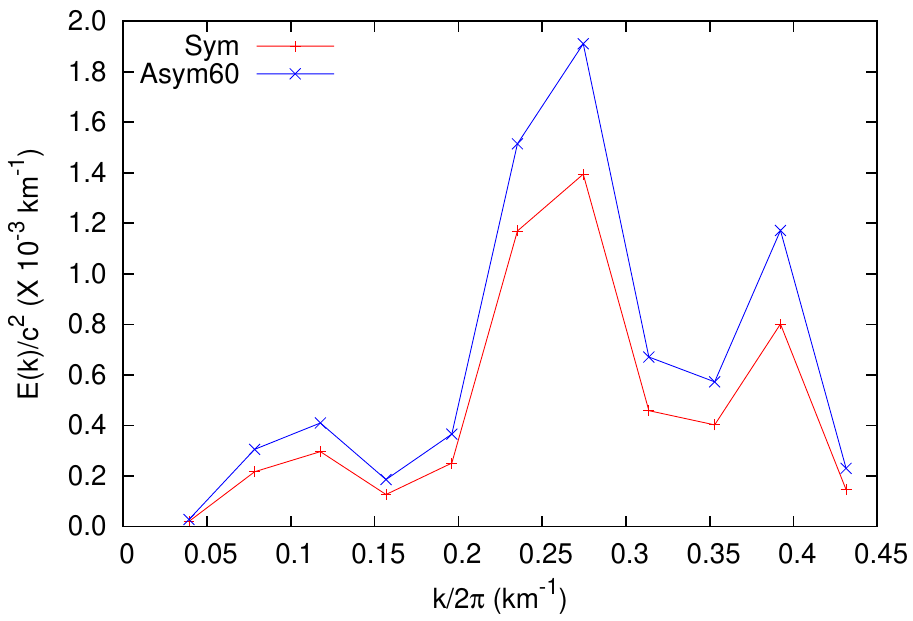}
\caption{
Comparison of the \textit{mean} specific kinetic energy spectra between {\bf Sym} and {\bf Asym60} at $t-t_\text{merge}\approx 30$ms.
}
\label{fig:kt30mean}
\end{center}
\end{figure}

For all the cases, a steady wind is formed along the disk edges abutting the corona close to the BH along the rotation axis. 
The outflow is only mildly relativistic, with the Lorentz factor ranging from 1.2 to 1.75.
We measure the amount of matter that becomes unbound by the criterion $u_t<-1$, where $u_t$ is the time component of the covariant fluid 4-velocity.
Measuring the change of this amount in time (Fig.~\ref{fig:unbound}), we see that the steady wind is launched at the same time as when the unbound matter 
starts to increase almost linearly from the $\approx 0.01M_{\odot}$ value achieved right after the merger.
The latter agrees with the early postmerger ejecta mass observed for BHNS systems where the NS is modeled with the APR4 EOS in Refs.~\cite{Hotoke13,Kyutoku15,Kawaguchi16}.

At the beginning of the second wind episode, we show the plasma density and the velocity field for the {\bf Asym60} model [Fig.~\ref{fig:wind}].
Simultaneous with the wind, we observe the evacuation of a funnel region with low $\beta$ above and below the equatorial plane [Fig.~\ref{fig:plasmabet}],
where $\beta$ is the fluid-to-magnetic energy density ratio defined as $2\rho/b^2$, with $\rho$ and $b^2={b_{\mu}}^{\mu}$ with the same definitions as for Eq.~\ref{eq:lambdaMRI}.
For all the cases, the Poynting flux measured at a distance of $\approx 960$km from the center of the system is found to climb up by about 5 orders of magnitude
during the beginning of the first wind episode. Figure~\ref{fig:poynflux} shows this for the asymmetric cases.
At the end of the first wind episode, the Poynting flux is increased up to an almost stable value of $\approx 2\times10^{49}$ erg/s.

An investigation on the thermal component of the specific internal energy reveals the formation of sharp thermal shock fronts 
in the wind regions, indicating that the outflow is highly thermalized (Fig.~\ref{fig:therm}). 
On the other hand, the ratio of the magnetic energy to the plasma kinetic energy for all the cases is measured to be at most $\approx 0.2\%$ even after the onset of the second wind episode (Fig.~\ref{fig:EBEkin}). 
This suggests that the wind episodes are not likely to be powered by magnetic energy.
Similar to the discussion given in Ref.~\cite{Kiuchi15}, we conjecture here that, it is the thermalization that leads to pressure gradients pushing the matter upward from the equator to the low $\beta$ regions.
Subsequently, the matter becomes unbound when it enters the region where $u_t<-1$.

The magnetic field, which is frozen in with the matter in our simulations, is dragged along with the wind outflow, forming large scales along the poloidal direction. A degree of coherence in the field emerges over the length scale of the disk edges abutting the corona, i.e., over a few tens of km. We show close-up snapshots of this coherence in the form of the magnetic vector field on the meridional plane in the disk region around the BH (Fig.~\ref{fig:coherentclose}). Coinciding with this coherence along the disk edges, the field is highly turbulent \textit{inside} the accretion disk as expected given the turbulent plasma flow. 

Looking further into the thermalized regions, we obtain the specific kinetic energy spectra in these volumes of interest.
Following Ref.~\cite{Kiuchi15}, the specific \textit{turbulent} kinetic energy spectrum is computed as a correlation of the velocity perturbations as follows:
\begin{equation}
E(k)=\frac{1}{2}\int\int_V\sum_j e^{i\vec{k}.\vec{r}} \delta v^j(\vec{x}+\vec{r})\delta v^j(\vec{x}) d^3rd\Omega_k,
\end{equation}
with $\vec{k}$ denoting the wave number vector, $k=\|\vec{k}\|$, $d\Omega_k$ the volume element in a spherical shell between $k$ and $d+dk$, and $V$ a cubic domain.
The turbulent kinetic energy spectra are obtained as the \textit{mean} of the spectra in eight cubic domains. Four domains are taken in the upper hemisphere of the system: firstly, a cube spanned by $x\in[25,75]$km, $y\in[-25,25]$km, and $z\in[25,75]$km; secondly, one spanned by $x\in[-75,-25]$km, $y\in[-25,25]$km, and $z\in[25,75]$km; thirdly, one spanned by $x\in[-25,25]$km, $y\in[-75,-25]$km, and $z\in[25,75]$km; and fourthly, a cube spanned by $x\in[-25,25]$km, $y\in[25,75]$km, and $z\in[25,75]$km. Four similar cubic domains are taken in the lower hemisphere, where they have the same $x$ and $y$ spans but with $z\in[-75,-25]$km.
We take the \textit{mean} spectra to be representative of the \textit{overall} turbulent kinetic energy in the turbulent regions.
Deviations from the \textit{mean} reflect the highly asymmetric configuration of the plasma flow around the BH as well as the magnetic field embedded inside, as could be expected from a merger scenario.
$\vec{x}$ for the velocity perturbations is taken to be the position vector for the center of the cubic domain, whereas $\vec{r}$ is taken to be the position vector relative to this center.
The velocity perturbations, $\delta v^i$, are evaluated as $v^i-\langle v^i\rangle$, where $\langle v^i\rangle$ is the volume average of the velocities in the respective cubes. We then take the wave number of the mode which generates the highest amplitude in the kinetic energy spectrum and denote it as $k_\text{max}$.
Figure~\ref{fig:kmaxt} shows the change in time of the \textit{mean} kinetic energy of this dominant mode.
The trend that is apparent from this measure is that the \textit{mean} kinetic energy in all the thermalized volumes is decreasing 
even as the episodes of steady wind are occurring. 
Assuming that the mechanism described in Ref.~\cite{Kiuchi15} holds here, a decreasing mean kinetic energy in time would mean a decreasing efficiency in the conversion of mass accretion energy into thermal energy as time progresses. This appears to be consistent with the expectation that the ejected mass would asymptote to a given amount at late times. 

At $t-t_\text{merge}\approx 30$ms which coincides with the beginning of the first wind episode, we compare the turbulent kinetic energy spectra between the symmetric {\bf Sym} and asymmetric case {\bf Asym60} (Fig.~\ref{fig:kt30mean}).
The \textit{mean} amplitude of the spectra for the symmetric case is decidedly lower than that for the asymmetric one.
This could indicate that the helicity in the magnetic field is generating a higher \textit{overall} turbulent kinetic energy.
At $t-t_\text{merge}\approx 70$ms, the difference between the ejected masses between {\bf Sym} and {\bf Asym60} reaches $\approx 20\%$ of the total ejected mass for {\bf Sym}. 

In an unprecedented resolution study performed in Ref.~\cite{Kiuchi15}, it is seen that increasing resolution yields higher ejected mass at a given time.
As mentioned above, this has been explained as due to the higher efficiency with which the thermalization of the mass accretion energy takes place,
which is attributed to the higher \textit{effective} turbulent viscosities realized at higher resolutions.
The total amount of ejected mass from the highest resolution run in Ref.~\cite{Kiuchi15} in the time scale simulated is thus designated as the lower bound.
Since certain qualitative features of the nonaxisymmetric MRI appears to have been captured in all our cases, we hypothesize that a similar resolution study performed for the cases in the current paper would produce a similar trend,
i.e., that with increased resolution, the efficiency of the thermalization of the mass accretion energy would increase, 
leading to higher ejected mass at a given time, or in other words, an earlier mass ejection episode for \textit{all} the cases here, both symmetric and asymmetric.
The increase in ejected mass at a given time could reach $400\%$ when the resolution is doubled (see Fig. 2 of Ref.~\cite{Kiuchi15}), 
much higher than the $20\%$ increase that could be reached purely by a helicity-induced enhancement of the turbulent kinetic energy as seen in the current study.

It has to be noted that even in Ref.~\cite{Kiuchi15} where 32,768 CPUs on the K supercomputer are used in their highest resolution run,
convergence in the amount of ejected mass is still not seen.
Indeed, since the dissipation happens on the grid scale, convergence would not be seen with increasing resolution, unless a large explicit viscosity, resolvable by the given resolutions, is introduced \cite{Balbus98}. 
Nonetheless, confirming whether the total amount of ejected mass would increase with the \textit{degree} of asymmetry
would still call for a thorough resolution study as well as a systematic one, so as to gauge whether the above-mentioned increases in the ejected mass will be trumped out by finite-resolution errors.
Because of the prohibitive computational resources required, we are presently unable to present such studies.
Using the results presented in the current paper as hints, we intend to perform the latter once the problem of computational resources becomes tractable.

\section{\label{sec:level4}Summary}

In this study, we conduct long-term GRMHD simulations of BHNS mergers (up to $\approx 100$ms after merger) where the NS is tidally disrupted by the BH forming a massive accretion disk.
The NS is endowed with an asymmetric dipole with a helical component. We study three cases where one case is endowed with a symmetric dipole and two cases are endowed with two different degrees of helicity respectively.
The magnetic field amplification in the merger remnant is clarified by performing a linear analysis based on the nonaxisymmetric MRI \cite{Balbus92}, and comparing with the growth rates of the magnetic field in the accretion disk obtained from the simulations.  
Both the toroidal and poloidal fields are amplified exponentially by the MRI up to an order of magnitude.
The toroidal field follows linear theory very closely except for the case with the highest degree of helicity studied here whereas the poloidal field growth is largely nonlinear due to poor resolvability and turbulence.
The deviation of the latter from linear theory prediction increases as asymmetry increases, in line with the increase in turbulence in the asymmetric cases. 
The MRI growth is estimated to be in the ideal MHD regime throughout almost the entire bulk of the accretion disk, so would thus be free from effects induced in the event neutrino radiation occurs.
We conclude that our cases with the resolution employed have eluded the most rapidly growing modes for the nonaxisymmetric system, even though certain qualitative features of the nonaxisymmetric MRI, such as the sustained turbulence, have been captured.    
Two strong thermally driven wind episodes along a funnel wall above and below the BH close to its rotation axis are uncovered in all the cases.
The formation of a large-scale coherent field in the poloidal direction is accompanied by a strong Poynting flux, both of which could enhance the prospects of sGRB jet formation.
However, due to the relatively small variation in the amount of ejected mass, we are unable presently to confirm whether
the asymmetry in the initial magnetic field indeed produces enhanced mass ejection corresponding to the \textit{degree} of asymmetry without a thorough resolution study.

\begin{acknowledgments}
The simulations were performed on SR16000 at YITP of Kyoto University
and with supercomputing resources including technical support 
at the Korean National Institute of Supercomputing and Network (NISN).
The work was also supported by Chinese NSFC Grants No. 11375153 and No. 11675145, and by IBS through Project Code (Grant No. IBS-R024-D1).
MBW is grateful to Kenta Kiuchi and Masaru Shibata for providing the code for the simulations and for helpful discussions,
as well as to Koutarou Kyutoku for providing the initial data.
MBW wishes to acknowledge Zach Etienne for helpful comments, Anzhong Wang for proofreading, Youngman Kim and Gentaro Watanabe for the support in acquiring the computational resources at NISN, and Ki-Young Choi for the hospitality at KASI where part of this work was done.
\end{acknowledgments}

\end{document}